%% file: main.tex
\documentclass[sigconf]{acmart}
\usepackage{hyperref}
\AtBeginDocument{%
  }

\setcopyright{cc}
\setcctype{by}
\copyrightyear{2026}
\acmYear{2026}
\acmDOI{}
\acmConference[\href{https://agentic-arch.org/}{$A^3$ Workshop at MICRO 2026}; Hackathon Spotlight Paper]{\href{https://agentic-arch.org/}{$A^3$: Agentic Approaches to Architecture Workshop at MICRO 2026}; Hackathon Spotlight Paper}{Nov 01, 2026}{Athens, Greece}
\acmBooktitle{\href{https://agentic-arch.org/}{$A^3$: Agentic Approaches to Architecture Workshop at MICRO 2026}; Hackathon Spotlight Paper, Nov 01, 2026, Athens, Greece}
\acmISBN{}
\usepackage{algorithm}
\usepackage{algpseudocode}
\algtext*{EndWhile}\algtext*{EndIf}\algtext*{EndFor}
\usepackage[capitalise,nameinlink]{cleveref}
\crefname{figure}{Figure}{Figures}
\crefname{table}{Table}{Tables}
\crefname{algorithm}{Algorithm}{Algorithms}
\crefname{section}{Section}{Sections}

\input{generated/numbers_v2-final15.tex}

\input{generated/numbers_v3-dnn512.tex}

\input{generated/numbers_v3-generalization.tex}

\input{generated/numbers_v3-transfer.tex}

\input{generated/numbers_v3-repair-bench.tex}

\input{generated/numbers_v3-dnn512-design.tex}

\input{generated/numbers_v3-generalization-design.tex}

\input{generated/numbers_measurement-validation.tex}

\input{generated/numbers_v2-corrections.tex}
\input{generated/numbers_v3-dnn512-ppa.tex}

\input{generated/numbers_v3-generalization-ppa.tex}

\input{generated/numbers_v3-workloads.tex}

\input{generated/numbers_v3-config.tex}

\input{generated/numbers_v3-transfer-grid.tex}

\input{generated/numbers_v3-walltime.tex}

\newcommand{\x}{\ensuremath{\times}}
\newcommand{\stage}[1]{\textsf{#1}}

\begin{document}

\title{SparseCraft: Agentic Hardware--Software Co-Optimization for Sparse Computing}

\author{M P Samartha}
\authornote{These authors contributed equally to this research.}
\affiliation{%
  \institution{International Institute of Information Technology}
  \city{Hyderabad}
  \country{India}}
\email{m.samartha@students.iiit.ac.in}
\author{Rajatabha Chakraborty}
\authornotemark[1]
\affiliation{%
  \institution{International Institute of Information Technology}
  \city{Hyderabad}
  \country{India}}
\email{rajatabha.c@research.iiit.ac.in}
\author{Vedant Pahariya}
\authornotemark[1]
\affiliation{%
  \institution{International Institute of Information Technology}
  \city{Hyderabad}
  \country{India}}
\email{vedant.pahariya@research.iiit.ac.in}
\author{Priyesh Shukla}
\affiliation{%
  \institution{International Institute of Information Technology}
  \city{Hyderabad}
  \country{India}}
\email{priyesh.shukla@iiit.ac.in}
\renewcommand{\shortauthors}{Samartha et al.}
\newcommand{\eqstar}{\texorpdfstring{\textsuperscript{*}}{}}
\renewcommand{\authors}{M P Samartha\eqstar\and Rajatabha Chakraborty\eqstar\and
  Vedant Pahariya\eqstar\and Priyesh Shukla}

\begin{abstract}
Sparse-accelerator design spaces are usually searched against analytical models, so a
design is judged by a model's prediction rather than by a measurement of the
hardware. SparseCraft is a closed loop on the CHIA framework in which a large
language model (LLM) agent co-designs the configuration, memories and sparse-kernel
schedule of a Gemmini accelerator. A candidate counts only after it passes a legality check,
cycle-accurate simulation, a bit-for-bit comparison of every output and synthesis.
Post-synthesis timing analysis checks its clock, and its energy is computed from
measured counters, with per-event energies validated against gate-level and
SRAM-macro data. We compare the agent with expert Gemmini configurations and with
random, greedy and multi-objective TPE search at the agent's budget of
\vthreeXOneSFullEvals{} evaluations, all measured by the same flow. On a
GraphChallenge sparse-DNN layer, the agent's designs reach
\vthreepXOneSBestEdapEdapGain--\vthreepXOneABestEdapEdapGain\x{} the inverse
energy--delay--area product of stock Gemmini and
\vthreepXOneSBestEdapVsEThree--\vthreepXOneABestEdapVsEThree\x{} that of the best
expert configuration. Their Pareto hypervolume is \vthreeXOneSHv--\vthreeXOneAHv{},
against at most \vthreeXTwoHv{} for search. On a power-law graph and a pruned BERT
layer, the agent raises perf/W by \vthreegenXFiveWGainPerfW\x{} and
\vthreegenXFiveBGainPerfW\x, and held-out transfer shows that its designs specialise
to their workload. The loop is also resilient to faulty designs. When a design fails
a check, a repair agent diagnoses the failure from its evidence and fixes it
automatically, or declines when the only fix would undo the intended change. In a
fault-injection benchmark, it fixed every scorable fault at its first attempt.
\end{abstract}

\keywords{LLM agents, sparse accelerators, design-space exploration,
hardware/software co-design, Gemmini, CHIA}

\maketitle

\section{Introduction}
\label{sec:intro}

Large language models (LLMs) now write hardware description code that compiles and
passes simulation~\cite{liu2023verilogeval,blocklove2025autochip}, set accelerator
parameters~\cite{fu2023gpt4aigchip,wang2025llmdse} and optimise accelerator
kernels~\cite{hong2025autocomp}. Most of this work generates a design once or scores
it with an analytical model, and kernel optimisers tune a single layer. A co-design
loop is harder. The model must read a \emph{measured} outcome and change the RTL, the
memories or the software in response, over many iterations that build on its own
earlier changes. The evaluation itself must stay out of its reach.

Sparsity makes the problem cross-layer, because accelerators exploit zeros in
different places. Dedicated datapaths \emph{skip} them. Outer-product and merge-based SpGEMM
engines~\cite{pal2018outerspace,zhang2021gamma}, intersection units for sparse tensor
algebra~\cite{hegde2019extensor} and flexible reduction networks~\cite{qin2020sigma}
issue only nonzero work, at the price of a datapath built around sparsity. Dense
arrays can instead \emph{gate} zeros, which saves switching energy but no
cycles~\cite{chen2017eyerissjssc}. They can also exploit \emph{structure}: block and
N:M sparsity keep a dense datapath busy~\cite{jeong2023vegeta,wu2023highlight}, and
offline reordering packs scattered nonzeros into fewer tiles~\cite{ji2018tetris}.
Which technique pays off depends on the workload's structure and on costs that are
easy to misjudge. Gating and skipping save different things~\cite{wu2022sparseloop},
and a technique that a model credits can cost energy once measured
(\cref{sec:measure}).

We study this on Gemmini~\cite{genc2021gemmini}, a Chisel-generated
weight-stationary systolic array in Chipyard~\cite{amid2020chipyard}. Its
block-sparse kernel skips empty $16{\times}16$ tiles, but in our sparse-DNN layer
the occupied tiles are only \datWlInBlockPct\% dense, so \datWlIssuedMacs{} MACs
are issued for \datWlUsefulMacs{} useful ones.
Recovering that waste can take coordinated changes to the kernel's schedule, the
memory hierarchy and the RTL. Such combinations are tedious to search by hand, and
each candidate is expensive to measure.

\begin{figure*}[t]
  \centering
  \includegraphics[width=\textwidth]{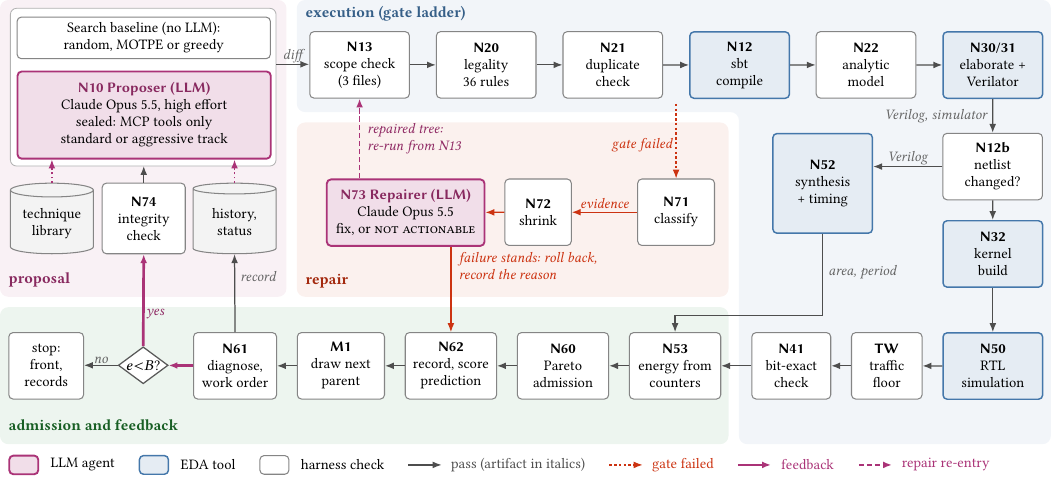}
  \caption{\textbf{The SparseCraft loop on CHIA.} Purple: LLM agents; blue: EDA tools
    in containers; white: harness checks; labels as in \cref{alg:loop}. Clockwise
    from the top left, a proposal source (the LLM proposer \stage{N10} or a search
    baseline) edits the design and feeds the gate ladder, where synthesis
    (\stage{N52}) runs beside simulation. A failed gate (red) goes to the repairer
    (\stage{N73}), whose fixed tree re-enters at \stage{N13}. \stage{N60} admits a
    measured design to the Pareto front or a MAP-Elites archive and refuses ties.
    \stage{M1} draws the next parent from both, and \stage{N61} writes the next work
    order (thick purple), until the number of full evaluations $e$ reaches the budget
    $B$.}
  \Description{Flow diagram of one iteration of the loop. A proposal source, an LLM
    agent or a search baseline, edits the design; a gate ladder of harness checks and
    tool runs (scope, legality, identity, compile, elaboration, kernel build,
    simulation, bit-exact check), with synthesis in parallel, measures it; a failed
    gate goes to a repair agent; the harness admits the design to a Pareto front or
    archive, draws the next parent, and writes the next work order.}
  \label{fig:loop}
\end{figure*}

SparseCraft is a closed loop for this search, built on CHIA~\cite{cui26}, an
open-source framework for agentic hardware/software co-design. The LLM is a node
inside the loop (\cref{fig:loop}). Each iteration begins with the LLM reading the
diagnosis, counters and history of the previous iteration, and ends with a measured
design that the harness admits or rejects. No human proposes, edits or selects a
design, and the harness, energy model and benchmark are fixed and hash-checked. Our
main question is whether an agent with measured feedback beats automated search and
expert configurations at the same small budget (\cref{sec:res-dnn}). We also test
whether its designs carry over to other kinds of sparsity or specialise
(\cref{sec:res-gen}), and whether its repair agent fixes broken designs and declines
when the only fix would undo the intended change (\cref{sec:res-repair}). Our
contributions are:
\begin{itemize}
  \item A closed loop in which a sealed LLM agent co-designs the configuration,
        memories and kernel schedule from measured feedback, in a standard and an
        aggressive track. A gate ladder builds, simulates, checks bit-for-bit and
        synthesises every proposal, and a repair agent handles failed gates
        (\cref{sec:method}).
  \item A measurement flow that checks every design's clock with post-synthesis
        timing and prices its energy from measured counters, with per-event energies
        validated against gate-level and SRAM-macro data. Its gate-level simulation
        shows that zero gating raises the energy per cycle of Gemmini's processing
        element on our workload (\cref{sec:measure}).
  \item An evaluation against expert configurations and random, greedy and
        multi-objective TPE search at an equal budget on structured sparsity, with
        agent runs on power-law and unstructured sparsity, held-out transfer and a
        fault-injection benchmark (\cref{sec:setup,sec:results}).
\end{itemize}

\section{The Closed Loop}
\label{sec:method}

\subsection{Structure}
\label{sec:structure}

\Cref{alg:loop} lists one run, and its stages are the nodes of \cref{fig:loop}.
Iteration~1 measures the initial design. Each iteration first checks that the harness,
scorer and benchmark still hash to their recorded values (\stage{N74}), and each run
uses a read-only snapshot of the harness. The budget counts \emph{full evaluations}:
proposals whose simulation ran, excluding the initial design. A run finishes when this
count reaches the set budget, independently of the model.

\begin{algorithm}[t]
  \caption{One SparseCraft run (stage labels as in \cref{fig:loop}).}
  \label{alg:loop}
  \footnotesize
  \begin{algorithmic}[1]
    \Require initial design $d_1$, evaluation budget $B$, repair budget $R$, parent rule $\pi$
    \State $p\gets d_1$; $\mathcal{F}\gets\emptyset$; $H\gets[\,]$; $\delta\gets\varnothing$; $Q\gets[\,]$; $e\gets 0$
    \For{$i=1,2,\dots$ \textbf{while} $e<B$} \Comment{$e$: full evaluations}
      \State \Call{AssertImmutable}{\,} \Comment{N74}
      \If{$i=1$} $\tau\gets d_1$
      \ElsIf{$Q\neq[\,]$} $\tau\gets$ \Call{Ablate}{$Q$} \Comment{aggressive only}
      \Else\ $(\tau,\hat y)\gets$ \Call{Propose}{$p,\delta,H$} \Comment{N10, LLM}
      \EndIf
      \State $v\gets$ \Call{Ladder}{$\tau$}; $r\gets 0$ \Comment{N13--N41}
      \While{$v$ failed \textbf{and} $v$ repairable \textbf{and} $r<R$}
        \State $x\gets$ \Call{Evidence}{$v$} \Comment{N71, N72}
        \State $(\tau,s)\gets$ \Call{Repair}{$x,p,\tau$}; $r\gets r{+}1$ \Comment{N73, LLM}
        \If{$s=$ \textsc{not actionable} \textbf{or} revert} \textbf{break}
        \EndIf
        \State $v\gets$ \Call{Ladder}{$\tau$}
      \EndWhile
      \If{$i>1$ \textbf{and} $v$ reached simulation} $e\gets e{+}1$
      \EndIf
      \If{$v$ failed} \State record $(\tau,v)$ in $H$; $\hat y$ unscorable
      \Else
        \State $(t,E,A)\gets$ \Call{Measure}{$v$} \Comment{N52, N53}
        \State $a\gets$ \Call{Admit}{$\mathcal{F},(t,E,A)$} \Comment{N60, ties refused}
        \State score $\hat y$ against $p$'s measurement; record $(\tau,a)$ in $H$ \Comment{N62}
        \If{$a$ admitted \textbf{and} $\tau$ moved several fields} queue ablations in $Q$
        \EndIf
      \EndIf
      \State $p\gets\pi(\mathcal{F})$; restore $p$'s exact tree \Comment{parent draw}
      \State $\delta\gets$ \Call{Diagnose}{$v$} \Comment{N61}
    \EndFor
    \State \Return $\mathcal{F},H$
  \end{algorithmic}
\end{algorithm}

\subsection{Agents and tracks}
\label{sec:proposer}

The proposer (\stage{N10}) is Claude Opus~5.5~\cite{anthropic2026opus55} at high
reasoning effort. It is sealed and acts only through Model Context Protocol (MCP)
tools served by CHIA. These are an editing
shell in the build container, status and history tools that list every design tried
with its result, and a frozen library of cards on published sparsity techniques. The
writable files hold the design parameters, the processing element (PE) and the
zero-bitmap unit (ZBU); \stage{N13} rejects an edit to any other file. Each work
order gives the parent's state, its diagnosis and counters, strategy modules for the
bottleneck and the candidates not yet tried.

The \emph{standard} track proposes from a short candidate list. The \emph{aggressive}
track must list more candidates under per-category quotas, and it loses a random
subset of its strategy modules each iteration to push it away from familiar moves. It
also spends budget on leave-one-out ablations of admitted multi-field changes, so
that a lucky combination cannot hide which field helped. The repair agent
(\stage{N73}) is the same model with a debugging method and a confidence protocol. It
fixes only a root cause it can name; otherwise it reports \textsc{not actionable},
and the stated cause goes into the next work order.

\subsection{Design space and legality}
\label{sec:legality}

A design is a typed record of hardware and schedule fields. The hardware fields set
the array geometry, memory capacities and banking, DMA and queue depths, and whether
the normalisation unit is built. The schedule fields set dense-operand residency,
move-in widths, offline row reordering of the static sparse
operand~\cite{ji2018tetris}, per-tile trimming and a stationary-operand order. The
harness also supplies reference RTL for a zero-gated MAC~\cite{chen2017eyerissjssc}
and for the ZBU, which suppresses scratchpad reads of all-zero
rows~\cite{wu2022sparseloop}; the agent can switch either on and rewrite its Chisel
logic. Before anything is built, legality (\stage{N20}) checks named rules: generator
invariants, hardware--kernel couplings, a metadata budget, and an array dimension
equal to the workload's block size, the only size at which the kernel computes
correct outputs. \stage{N21} rejects a design whose state and RTL digest match one
already evaluated.

\subsection{Measurement}
\label{sec:measure}

Elaboration and a Verilator~\cite{snyder2023verilator} build (\stage{N30/31}) give a
cycle-accurate simulator of a Rocket core with Gemmini. The kernel (\stage{N32})
walks only the nonzero $16{\times}16$ blocks, and simulation (\stage{N50}) yields
cycles, off-chip bytes and Gemmini's counters. A simulation is rejected if its
off-chip traffic is implausibly low, below a quarter of the bytes needed to read its
inputs and write its output once. \stage{N41} then compares every output with a
golden result computed on the host. In parallel,
synthesis (\stage{N52}) maps the design to NanGate45 cells~\cite{nangate45opencell}
with Yosys~\cite{wolf2016yosys} and tiles every SRAM bank with fakeram45
macros~\cite{bsg2023fakeram}. After a fan-out buffering pass in
OpenROAD~\cite{ajayi2019openroad}, OpenSTA~\cite{opensta} times the pre-placement
netlist with a wire-load model. Area is the sum of logic and SRAM-macro area.

\emph{Clock.} The period is the larger of the \cfgClockFloorNs\,ns floor and the
static-timing period of the logic the agent can change (mesh, PE and ZBU), so an edit
that slows this logic pays with a slower clock. The longest path in the whole tile
(\measBaseTileArrivalNs\,ns) lies in Gemmini's floating-point accumulator-scale unit.
Its pipeline registers would shorten this path only if synthesis retimed them, and
Yosys does not retime. No design changes this unit, so we record its path but do not
use it to set the clock. The stock mesh
path arrives at \measBaseMeshArrivalNs\,ns, and the scoped logic of every reported
design also meets the floor, so all of them run at \cfgClockFloorNs\,ns.

\emph{Energy} (\stage{N53}) is computed from the measured counters. MAC energy
multiplies the issued MACs by a per-MAC energy from a gate-level simulation of the
generated PE on dnn512's operand stream. We apply this cost to every workload; MAC
energy is at most \vthreegendXFiveBBestPerfWMacPct\% of any best design's energy. On
this stream, zero gating raises the PE's energy per cycle from \measPeUngatedPJ{} to
\measPeGatedPJ\,pJ (+\measPeGatingPct\%). SRAM energy prices the kernel's command
stream at fakeram45 per-access energies and is a small share of every design's
energy. DRAM energy prices the measured off-chip bytes at a GDDR5-class
\measSensGddrPJB\,pJ/B~\cite{oconnor2017fgdram}, and leakage of cells and macros
accrues over the run time.

\subsection{Admission, parent selection and repair}
\label{sec:admit}

\stage{N60} admits a measured design only if no admitted design matches or beats it
in time, energy and area, so a design that ties one already held is refused. A
MAP-Elites archive also keeps designs that fill an empty niche, as stepping stones.
After every iteration, a seeded rule (\stage{M1}) draws the next parent from the
front and the archive, weighting each design by its measured quality and by how few
proposals have already started from it. \stage{N61} labels the bottleneck from the
DMA-wait and array-activity counters and selects the next strategy modules.
\stage{N62} records every proposal together with the agent's \emph{prediction} for
each objective (better, flat or worse than the parent). It scores the prediction if
the design was measured and marks it unscorable otherwise. Infrastructure faults
never reach a model (\stage{N71}). The repairer receives the shrunk evidence
(\stage{N72}), including a failed build's root exception, both design states and its
earlier attempts. It returns either a fixed tree, which re-runs the ladder, or
\textsc{not actionable}. Reverting repairs are detected, and attempts are capped.

\begin{table}[t]
  \caption{Workloads; blocks: the occupied $16{\times}16$ tiles of $A$.}
  \label{tab:workloads}
  \footnotesize
  \setlength{\tabcolsep}{3.5pt}
  \begin{tabular}{@{}lrrrrl@{}}
    \toprule
    workload & $A$ & $N$ & nnz (density) & blocks & source \\
    \midrule
    dnn512      & $\wlDnnRows{\times}\wlDnnCols$ & \wlDnnDenseCols & \wlDnnNnz{} (\wlDnnDensityPct\%) & \wlDnnNzBlocks{} of \wlDnnTotalBlocks & GraphChallenge \\
    wikivote384 & $\wlWikiRows{\times}\wlWikiCols$ & \wlWikiDenseCols & \wlWikiNnz{} (\wlWikiDensityPct\%) & \wlWikiNzBlocks{} of \wlWikiTotalBlocks & SNAP wiki-Vote \\
    obert256    & $\wlObertRows{\times}\wlObertCols$ & \wlObertDenseCols & \wlObertNnz{} (\wlObertDensityPct\%) & \wlObertNzBlocks{} of \wlObertTotalBlocks & oBERT, pruned \\
    \bottomrule
  \end{tabular}
\end{table}

\section{Experimental Setup}
\label{sec:setup}

\subsection{Workloads}
\label{sec:workloads}

Every workload is $Y{=}AX$ with a static sparse $A$ and a dense INT8 $X$
(\cref{tab:workloads}); together they span structured, power-law and unstructured
sparsity. \emph{dnn512} is the upper-left block of the first layer of the
GraphChallenge sparse DNN~\cite{kepner2019graphchallenge}. Its pattern is structured,
so a static permutation packs its nonzeros into far fewer tiles. \emph{wikivote384}
is a tile of the SNAP wiki-Vote graph~\cite{leskovec2014snap}, a power-law matrix
from the SpGEMM set of OuterSPACE and Gamma~\cite{pal2018outerspace,zhang2021gamma}.
\emph{obert256} is a slice of a \cfgObertPrunePct\%-pruned BERT-base
layer~\cite{kurtic2022obert}, unstructured and denser. We use only the sparsity
patterns and fill them with seeded values that are never zero, which keeps the
bit-exact check sensitive to indexing errors.

\begin{table}[t]
  \caption{Best value of each column per arm on dnn512 within \vthreeXOneSFullEvals{}
    full evaluations. Columns may come from different designs; each expert tier is a
    single design. $1/\mathrm{EDAP}$ is relative to stock; evals to E3 counts the
    evaluations until the arm first matched E3's perf/W. The agents lead on perf/W,
    $1/\mathrm{EDAP}$ and HV, and their fastest designs need slightly fewer cycles than
    E3. Bold: the best arm in each column.}
  \label{tab:dnn}
  \footnotesize
  \setlength{\tabcolsep}{2.6pt}
  \begin{tabular}{@{}lrrrrr@{}}
    \toprule
    arm & GOPS/W & $1/\mathrm{EDAP}$ & cycles & HV & evals to E3 \\
    \midrule
    stock                 & \vthreeXOneSBasePerfW & $1$ & \vthreeXOneSBaseCycles & n/a & n/a \\
    E1 expert             & \vthreeEOneBestPerfW & \vthreepEOneBestEdapEdapGain & \vthreeEOneBestCycles & n/a & n/a \\
    E2 expert             & \vthreeETwoBestPerfW & \vthreepETwoBestEdapEdapGain & \vthreeETwoBestCycles & n/a & n/a \\
    E3 expert             & \vthreeEThreeBestPerfW & \vthreepEThreeBestEdapEdapGain & \vthreeEThreeBestCycles & n/a & n/a \\
    \midrule
    random                & \vthreeXThreeBestPerfW & \vthreepXThreeBestEdapEdapGain & \vthreeXThreeBestCycles & \vthreeXThreeHv & \vthreeXThreeEvalsToEThreePerfW \\
    greedy                & \vthreeXNineBestPerfW & \vthreepXNineBestEdapEdapGain & \vthreeXNineBestCycles & \vthreeXNineHv & \vthreeXNineEvalsToEThreePerfW \\
    MOTPE                 & \vthreeXTwoBestPerfW & \vthreepXTwoBestEdapEdapGain & \vthreeXTwoBestCycles & \vthreeXTwoHv & \vthreeXTwoEvalsToEThreePerfW \\
    \midrule
    agent, standard       & \vthreeXOneSBestPerfW & \vthreepXOneSBestEdapEdapGain & \vthreeXOneSBestCycles & \vthreeXOneSHv & \textbf{\vthreeXOneSEvalsToEThreePerfW} \\
    agent, aggressive     & \textbf{\vthreeXOneABestPerfW} & \textbf{\vthreepXOneABestEdapEdapGain} & \textbf{\vthreeXOneABestCycles} & \textbf{\vthreeXOneAHv} & \vthreeXOneAEvalsToEThreePerfW \\
    \bottomrule
  \end{tabular}
\end{table}

\subsection{Baselines}
\label{sec:baselines}

\emph{Stock.} Iteration~1 of every agent and search run is stock Gemmini
($16{\times}16$ weight-stationary array) running the block-sparse kernel, so every
gain is on top of skipping empty tiles.

\emph{Expert configurations.} A stronger reference is what an architect
would set by hand from published practice. Our expert tiers were frozen before any
measurement, and each builds on the one before. E1
keeps the dense operand resident on chip. E2 adds zero gating of the
MACs~\cite{chen2017eyerissjssc}, the ZBU~\cite{wu2022sparseloop} and memories sized
to the working set. E3 adds static row reordering~\cite{ji2018tetris}. Each tier is
measured by the same flow as every agent design, so a comparison isolates the quality
of the configuration. The tiers are not published sparse accelerators: designs such
as SIGMA or ExTensor use datapaths outside Gemmini's template and cannot be measured
in this flow.

\emph{Automated search.} At an equal budget of \vthreeXOneSFullEvals{} full
evaluations, we run random search, greedy coordinate search and MOTPE. Random search
draws every field uniformly. Greedy search changes one field at a time from an
incumbent, which moves to every design admitted to the front. MOTPE is Optuna's
multi-objective tree-structured Parzen estimator~\cite{ozaki2020motpe} over time,
energy and area. The searches cover the agent's typed space, kernel options included,
except a few generator values that do not elaborate; the agent never received this
filter. Each search is told every result. The agent's prompts guide it toward informed
decisions. They give each workload's block counts and the measured costs of zero
gating and of the kernel options, and they note that a resident dense operand and a
smaller scratchpad pay off. The expert configurations draw on the same knowledge but
are not part of the prompts.

\emph{Agent arms.} The standard and aggressive tracks (\cref{sec:proposer}) run on
dnn512 at the same budget, and the standard track also runs on wikivote384 and
obert256. Each arm is a single run, so we report no variance across seeds.

\subsection{Metrics}
\label{sec:metrics}

For a design, $t$ is the run time (cycles times the clock period), $E$ the energy,
$A$ the area and $U$ the number of useful operations ($2N$ per nonzero of the sparse
operand). We report performance ($U/t$), power ($E/t$), perf/W ($U/E$) and perf/area
($U/(tA)$). Energy and delay combine in the energy--delay product
($\mathrm{EDP}{=}Et$) and, with area, in the energy--delay--area product
($\mathrm{EDAP}{=}EtA$). Our single figure of merit is $1/\mathrm{EDAP}$ relative
to the stock design of the same workload (higher is better), so speed bought with
area or energy does not count as a gain. For each run, we also report the hypervolume
(HV) of its measured $(t,E,A)$ front, normalised to stock with reference point
\cfgHvRef. HV credits trade-offs that no single metric captures.

\section{Results}
\label{sec:results}

\begin{figure}[t]
  \centering
  \includegraphics[width=\columnwidth]{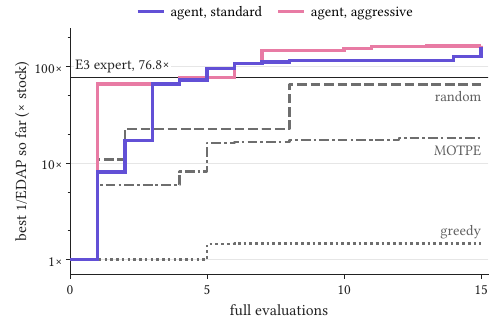}
  \caption{\textbf{Search on dnn512.} Best $1/\mathrm{EDAP}$ so far over stock (log
    scale) against full evaluations. Solid: agent tracks; broken grey: search
    baselines; black: E3. Both agent tracks pass E3 early and are still improving at
    the end, at \vthreepXOneSBestEdapEdapGain\x{} and
    \vthreepXOneABestEdapEdapGain\x{} stock. Random, MOTPE and greedy search end at
    \vthreepXThreeBestEdapEdapGain\x, \vthreepXTwoBestEdapEdapGain\x{} and
    \vthreepXNineBestEdapEdapGain\x.}
  \Description{Best-so-far inverse EDAP against evaluations for every arm.}
  \label{fig:search}
\end{figure}

\begin{figure*}[t]
  \centering
  \includegraphics[width=\textwidth]{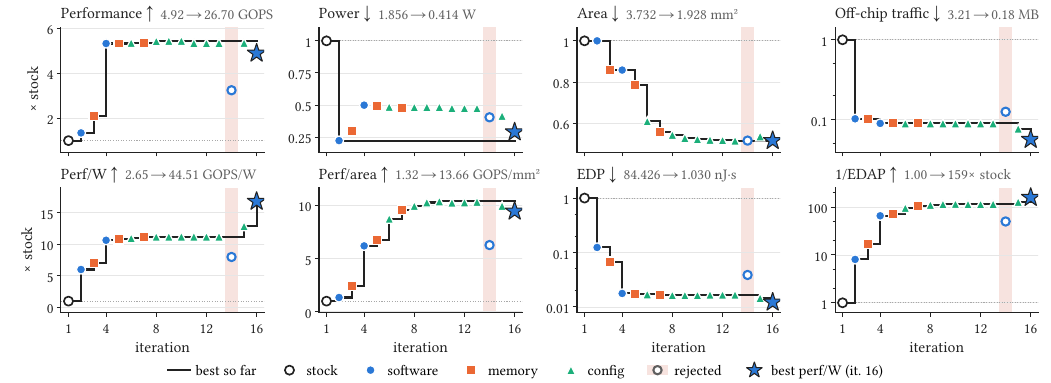}
  \caption{\textbf{Metric evolution of the standard agent track on dnn512},
    normalised to stock (iteration~1). Panel titles give the stock and best absolute
    values; off-chip traffic, EDP and $1/\mathrm{EDAP}$ use log scales. Markers:
    measured designs by kind of change (open and shaded: rejected); black step: best
    so far; star: best perf/W. Residency, a halved scratchpad and row reordering
    (it.~2 to~4) give most of the speed and energy gain, and later trims shrink the
    area. A narrower move-in of the dense operand (it.~\vthreedXOneSBestPerfWIter)
    cuts off-chip traffic and gives the best perf/W and $1/\mathrm{EDAP}$.}
  \Description{Small multiples of performance, power, area, off-chip traffic,
    perf/W, perf/area, EDP and inverse EDAP over the iterations of the standard track.}
  \label{fig:metrics}
\end{figure*}

\subsection{Agent versus expert and search on dnn512}
\label{sec:res-dnn}

\Cref{tab:dnn} and \cref{fig:search} compare the arms. The agent tracks lead on
perf/W with \vthreeXOneSBestPerfW{} and \vthreeXOneABestPerfW{} GOPS/W
(\vthreeXOneSGainPerfW\x{} and \vthreeXOneAGainPerfW\x{} stock), against
\vthreeEThreeBestPerfW{} for the best expert configuration and
\vthreeXThreeBestPerfW{} for the best search. Their best designs reach
\vthreepXOneSBestEdapEdapGain\x{} and \vthreepXOneABestEdapEdapGain\x{} the stock
$1/\mathrm{EDAP}$, which is \vthreepXOneSBestEdapVsEThree\x{} and
\vthreepXOneABestEdapVsEThree\x{} that of E3. The agents match E3's perf/W after
\vthreeXOneSEvalsToEThreePerfW{} and \vthreeXOneAEvalsToEThreePerfW{} evaluations.
Random search needs \vthreeXThreeEvalsToEThreePerfW, and MOTPE and greedy search
never get there; on $1/\mathrm{EDAP}$, no search baseline reaches E3 at all. The
agents' fronts have a hypervolume of \vthreeXOneSHv{} and \vthreeXOneAHv, against at
most \vthreeXTwoHv{} for search. The arms are closest on speed. The agents' fastest
designs take \vthreeXOneABestCycles{} to \vthreeXOneSBestCycles{} cycles against E3's
\vthreeEThreeBestCycles, and their best perf/W designs are slower than E3
(\vthreedXOneSBestPerfWCycles{} cycles for the standard track). Their lead therefore
comes from energy and area (\cref{sec:res-traj}).

The baselines fall short for different reasons. Random search finds a good schedule
but keeps the stock memories (\vthreedXThreeBestPerfWSramAreaMM{} against
\vthreedXOneSBestPerfWSramAreaMM\,mm$^2$ of SRAM for the standard track). MOTPE
samples at random until it holds \cfgMotpeStartup{} measured designs, stock included,
and its best perf/W design runs at \vthreedXTwoBestPerfWCycles{} cycles, close to
stock. Greedy search's seeded field order spent the budget on bank counts, DMA width
and queue depths and never reached residency, memory sizing or reordering
(\vthreeXNineGainPerfW\x{} stock perf/W).

The aggressive track ran leave-one-out ablations on its first multi-field change.
Without row reordering, the design was much slower and was rejected. Restoring the stock scratchpad made it slightly faster at a cost in area and
energy, and the front admitted it as a trade-off.

\subsection{Trajectory: what the agent changed}
\label{sec:res-traj}

\Cref{fig:metrics} follows the standard track. Its first move was a software change
that its prompts suggest: keeping all of the dense operand resident in the scratchpad
(iteration~2) removed most of the off-chip traffic, which carries most of the energy.
Halving the scratchpad (iteration~3) and reordering the sparse operand's rows so that
its nonzeros fill fewer tiles (iteration~4) brought the cycle count close to its
final value. Admissions through iteration~13 then trimmed area at little or no cost
in time. They cut the scratchpad and accumulator to quarter size, removed the
normalisation unit (this workload never uses it), shortened the scratchpad read delay
and made the request, queue and reservation-station depths shallower. A
stationary-operand ordering was rejected for its extra cycles (iteration~14), and a
wider DMA was admitted for its lower energy (iteration~15). The parent rule then
returned to iteration~13, and from there moving the dense operand in one tile wide
(iteration~\vthreedXOneSBestPerfWIter) cut DRAM energy and gave the best perf/W. That
design differs from stock in \vthreedXOneSBestPerfWNChanged{} fields. DRAM still
takes \vthreedXOneSBestPerfWDramPct\% of its energy (\vthreedXOneSBaseDramPct\% at
stock), and leakage takes \vthreedXOneSBestPerfWLeakPct\%.

\begin{figure}[t]
  \centering
  \includegraphics[width=\columnwidth]{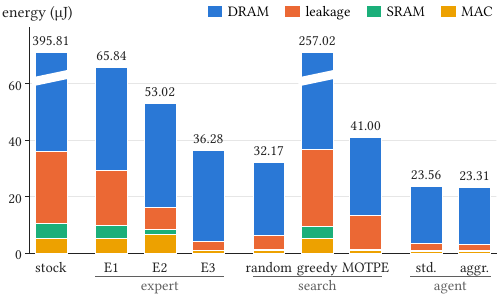}
  \caption{\textbf{Energy by source on dnn512.} Each bar is an arm's best perf/W
    design, which on a fixed workload is its lowest-energy design (std., aggr.: agent
    tracks). The stock and greedy bars are cut, with totals above. The agents spend
    the least on DRAM.}
  \Description{Stacked bars of MAC, SRAM, leakage and DRAM energy for the stock
    design, the expert tiers E1 to E3, random, greedy and MOTPE search, and the
    standard and aggressive agent tracks.}
  \label{fig:energy}
\end{figure}

The best designs of both dnn512 tracks share the same core: the resident dense
operand, row reordering, no normalisation unit and memories sized to the working set.
Every change the agents made is a generator parameter or a kernel option. Most of
their lead over E3 comes from levers that no expert tier uses. Removing the
normalisation unit and making the queues shallower cut logic area from
\vthreedEThreeBestPerfWLogicAreaMM{} to \vthreedXOneSBestPerfWLogicAreaMM\,mm$^2$, and
the narrower move-in cut DRAM energy.

\Cref{fig:energy} splits the energy of each arm's best perf/W design by source. DRAM
dominates every design. Keeping the dense operand resident cuts DRAM energy from
\vthreedXOneSBaseDramUJ{} to \vthreedEOneBestPerfWDramUJ\,$\mu$J (E1), and the
agents' narrower move-in brings it down to \vthreedXOneSBestPerfWDramUJ\,$\mu$J,
against \vthreedEThreeBestPerfWDramUJ\,$\mu$J for E3. Greedy search never makes the
operand resident, so its design still spends \vthreedXNineBestPerfWDramUJ\,$\mu$J on
DRAM. Shorter runs on smaller memories also cut leakage. MOTPE's design runs close to
stock speed and leaks \vthreedXTwoBestPerfWLeakUJ\,$\mu$J. Zero gating
raises E2's MAC energy above E1's (\vthreedETwoBestPerfWMacUJ{} against
\vthreedEOneBestPerfWMacUJ\,$\mu$J).

\subsection{Generalisation and held-out transfer}
\label{sec:res-gen}

\begin{figure}[t]
  \centering
  \includegraphics[width=\columnwidth]{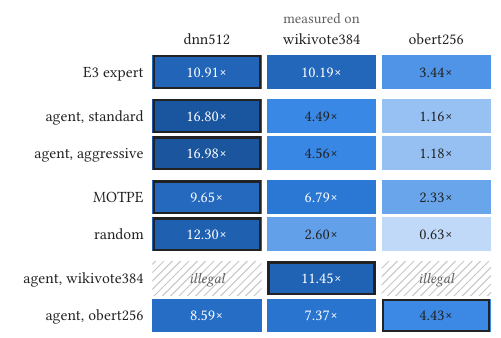}
  \caption{\textbf{Held-out transfer.} Perf/W gain of each arm's best design over each
    workload's stock design (darker: larger). Outlined: found on that workload;
    hatched: illegal (the dense operand no longer fits the scratchpad). In each
    column, the agent design found on that workload is best. On obert256, the dnn512
    agent designs keep only \vthreexgXOneSOnObertPerfWGain\x{} and
    \vthreexgXOneAOnObertPerfWGain\x, against \vthreexgEThreeOnObertPerfWGain\x{} for
    E3.}
  \Description{Heatmap of designs against workloads.}
  \label{fig:transfer}
\end{figure}

On the other workloads, the standard track raises perf/W by
\vthreegenXFiveWGainPerfW\x{} (wikivote384) and \vthreegenXFiveBGainPerfW\x{}
(obert256), and $1/\mathrm{EDAP}$ by \vthreegenpXFiveWBestEdapEdapGain\x{} and
\vthreegenpXFiveBBestEdapEdapGain\x. It uses the same core levers in different
proportions. On obert256, where every tile already holds a nonzero
(\wlObertNzBlockPct\%), it never reorders rows and widens the scratchpad banking
instead. \Cref{fig:transfer} re-measures each arm's best design, without retuning, on
the workloads it was not tuned for. On each workload, the design found there wins,
and every transferred design falls below it. On obert256, the dnn512 standard design
falls to \vthreexXOneSOnObertBestPerfW{} GOPS/W, barely above the stock design's
\vthreegenXFiveBBasePerfW, and runs slower than stock
(\vthreexXOneSOnObertBestCycles{} against \vthreegenXFiveBBaseCycles{} cycles). E3,
with the same memory sizes, keeps \vthreexEThreeOnObertBestPerfW{} GOPS/W there. The
wikivote384 design is illegal on the other workloads, because its scratchpad cannot
hold their dense operand. The agent's designs are therefore specialised to the
workload they were searched on, and each workload class needs its own search.

\subsection{Repair under fault injection}
\label{sec:res-repair}

No agent proposal in the runs above failed a gate, so the repair agent was never
called. To exercise it, we injected faults into known-good designs
(\cref{tab:repair}). Each scorable fault adds one defect to a correct change of the
PE, the ZBU or the schedule. The defects cover compile and elaboration errors,
functional bugs that produce wrong outputs, a hang, an RTL change that never reaches
the hardware, an edit outside the writable set and an illegal schedule. The repair
agent fixed \xsixFixed{} of \xsixPositiveScored{} scorable faults, each at its first
attempt, for \xsixCostUSD{} USD in total, and \xsixTruthMatches{} of the
\xsixTruthComparable{} comparable fixes reproduce the reference RTL exactly. It also
declined a negative control whose only repair would undo the intended change: a finer
zero-bitmap granule, whose bitmap exceeds the metadata budget. A second control, a
$32{\times}32$ array, is refused by legality before any repair is attempted. One
fault was excluded because its premise does not hold on this toolchain.

\begin{table}[t]
  \caption{Repair benchmark: faults injected into known-good designs. Fixed: the
    repaired design passes every gate; negative controls must be declined.}
  \label{tab:repair}
  \footnotesize
  \setlength{\tabcolsep}{3pt}
  \begin{tabular}{@{}llll@{}}
    \toprule
    class & injected failure & faults & outcome \\
    \midrule
    compile      & Chisel type errors, a renamed port      & C2, C3     & \xsixCTwoOutcome, \xsixCThreeOutcome \\
    elaboration  & ZBU output not driven                    & E1         & \xsixEOneOutcome \\
    divergence   & wrong mux polarity, stage, reduction     & F1--F3     & \xsixFOneOutcome{} (all) \\
    hang         & valid signal dropped                     & H1         & \xsixHOneOutcome \\
    RTL no-op    & toggle not set                           & N1         & \xsixNOneOutcome \\
    scope        & edit outside the writable set            & S1         & \xsixSOneOutcome \\
    legality     & schedule overflows the scratchpad        & T1         & \xsixTOneOutcome \\
    \midrule
    neg.\ control & ZBU bitmap over the metadata budget   & X1         & declined (\xsixXOneOutcome) \\
    neg.\ control & $32{\times}32$ array                     & X2         & refused by legality \\
    invalid      & premise false on this toolchain          & C1         & excluded \\
    \bottomrule
  \end{tabular}
\end{table}

\subsection{Cost and active time}
\label{sec:res-time}

We count only active time: each iteration's wall-clock time minus recorded lock and
memory waits. Time between iterations or runs is excluded, but queueing for shared
simulation slots is counted. An agent iteration on dnn512 took
\wallXOneSMeanIterMin{} minutes on average in the standard track and
\wallXOneAMeanIterMin{} in the aggressive track, with \wallXOneSLLMSharePct\% and
\wallXOneALLMSharePct\% of that time spent in the model. Model usage cost
\vthreeXOneSLLMCost{} and \vthreeXOneALLMCost{} USD per run.

\section{Conclusion}
\label{sec:conc}

SparseCraft puts an LLM agent inside a closed CHIA loop that co-designs the
configuration, memories and kernel schedule of a sparse Gemmini accelerator from
measured feedback. A design counts only after it is built, simulated, checked
bit-for-bit and synthesised. At an equal budget of \vthreeXOneSFullEvals{}
evaluations on a sparse-DNN layer, the agent's designs reach up to
\vthreeXOneABestPerfW{} GOPS/W and \vthreepXOneABestEdapEdapGain\x{} the stock
$1/\mathrm{EDAP}$, \vthreepXOneABestEdapVsEThree\x{} that of the best expert
configuration. Random, greedy and MOTPE search stay below the agents on both metrics.
On a power-law graph and a pruned BERT layer, the agent raises perf/W by
\vthreegenXFiveWGainPerfW\x{} and \vthreegenXFiveBGainPerfW\x, and its designs
specialise to their workload. The repair agent fixed every scorable injected fault.
The dnn512 agents were still improving when the budget ran out, so longer runs with
repeated seeds come next. We also plan a fan-out loop that expands several candidates
of a parent in parallel, so that promising alternatives are not lost to trial order.

\section{Artifact Information}
\label{sec:artifact}
The complete SparseCraft codebase and all the files required to replicate the work in
this paper are available in the repository\\
\mbox{\url{https://github.com/SARCS-IIITH/SparseCraft}}.

\begin{acks}
We thank Google for its support, the IIIT Hyderabad data centre for the computing
infrastructure used in this work, and the organizers of the CHIA Hackathon. Drafted with
AI assistance per the venue's policy; the authors own all claims.
\end{acks}

\clearpage
\balance
\bibliographystyle{ACM-Reference-Format}
\bibliography{paper}

\end{document}

%% file: generated/numbers_v2-final15.tex
\newcommand{\datWlInBlockPct}{6.25}
\newcommand{\datWlUsefulMacs}{524{,}288}

\newcommand{\datWlIssuedMacs}{8{,}388{,}608}

%% file: generated/numbers_v3-dnn512.tex
\newcommand{\vthreeXOneSFullEvals}{15}
\newcommand{\vthreeXOneSBaseCycles}{106{,}650}

\newcommand{\vthreeXOneSBasePerfW}{2.65}

\newcommand{\vthreeXOneSBestPerfW}{44.51}

\newcommand{\vthreeXOneSBestCycles}{19{,}638}

\newcommand{\vthreeXOneSGainPerfW}{16.80}
\newcommand{\vthreeXOneSHv}{0.5557}

\newcommand{\vthreeXOneSLLMCost}{8.12}

\newcommand{\vthreeXOneABestPerfW}{44.98}

\newcommand{\vthreeXOneABestCycles}{19{,}354}

\newcommand{\vthreeXOneAGainPerfW}{16.98}
\newcommand{\vthreeXOneAHv}{0.5704}

\newcommand{\vthreeXOneALLMCost}{6.70}

\newcommand{\vthreeXTwoBestPerfW}{25.58}

\newcommand{\vthreeXTwoBestCycles}{38{,}986}

\newcommand{\vthreeXTwoHv}{0.2806}

\newcommand{\vthreeXThreeBestPerfW}{32.59}

\newcommand{\vthreeXThreeBestCycles}{24{,}562}

\newcommand{\vthreeXThreeHv}{0.2612}

\newcommand{\vthreeEOneBestPerfW}{15.93}

\newcommand{\vthreeEOneBestCycles}{79{,}433}

\newcommand{\vthreeETwoBestPerfW}{19.78}

\newcommand{\vthreeETwoBestCycles}{50{,}862}

\newcommand{\vthreeEThreeBestPerfW}{28.90}

\newcommand{\vthreeEThreeBestCycles}{20{,}075}

\newcommand{\vthreeXNineBestPerfW}{4.08}

\newcommand{\vthreeXNineBestCycles}{106{,}650}

\newcommand{\vthreeXNineGainPerfW}{1.54}
\newcommand{\vthreeXNineHv}{0.0030}

\newcommand{\vthreeXOneSEvalsToEThreePerfW}{5}
\newcommand{\vthreeXOneAEvalsToEThreePerfW}{6}
\newcommand{\vthreeXTwoEvalsToEThreePerfW}{not reached}
\newcommand{\vthreeXThreeEvalsToEThreePerfW}{8}

\newcommand{\vthreeXNineEvalsToEThreePerfW}{not reached}

%% file: generated/numbers_v3-generalization.tex
\newcommand{\vthreegenXFiveWGainPerfW}{11.45}

\newcommand{\vthreegenXFiveBBaseCycles}{79{,}906}

\newcommand{\vthreegenXFiveBBasePerfW}{11.88}

\newcommand{\vthreegenXFiveBGainPerfW}{4.43}

%% file: generated/numbers_v3-transfer.tex
\newcommand{\vthreexEThreeOnObertBestPerfW}{40.93}

\newcommand{\vthreexXOneSOnObertBestPerfW}{13.83}

\newcommand{\vthreexXOneSOnObertBestCycles}{101{,}212}

%% file: generated/numbers_v3-repair-bench.tex
\newcommand{\xsixPositiveScored}{10}
\newcommand{\xsixFixed}{10}

\newcommand{\xsixCostUSD}{1.98}

\newcommand{\xsixTruthComparable}{9}
\newcommand{\xsixTruthMatches}{7}

\newcommand{\xsixFOneOutcome}{fixed}

\newcommand{\xsixNOneOutcome}{fixed}

\newcommand{\xsixSOneOutcome}{fixed}

\newcommand{\xsixHOneOutcome}{fixed}

\newcommand{\xsixCTwoOutcome}{fixed}

\newcommand{\xsixCThreeOutcome}{fixed}

\newcommand{\xsixEOneOutcome}{fixed}

\newcommand{\xsixTOneOutcome}{fixed}

\newcommand{\xsixXOneOutcome}{correct}

%% file: generated/numbers_v3-dnn512-design.tex
\newcommand{\vthreedXOneSBaseDramUJ}{359.66}

\newcommand{\vthreedXOneSBaseDramPct}{90.9}

\newcommand{\vthreedXOneSBestPerfWIter}{16}

\newcommand{\vthreedXOneSBestPerfWNChanged}{12}
\newcommand{\vthreedXOneSBestPerfWCycles}{21{,}859}

\newcommand{\vthreedXOneSBestPerfWDramUJ}{20.19}

\newcommand{\vthreedXOneSBestPerfWDramPct}{85.7}
\newcommand{\vthreedXOneSBestPerfWLeakPct}{10.1}

\newcommand{\vthreedXOneSBestPerfWLogicAreaMM}{1.586}
\newcommand{\vthreedXOneSBestPerfWSramAreaMM}{0.342}

\newcommand{\vthreedXTwoBestPerfWCycles}{97{,}435}

\newcommand{\vthreedXTwoBestPerfWLeakUJ}{12.18}

\newcommand{\vthreedXThreeBestPerfWSramAreaMM}{1.344}

\newcommand{\vthreedEOneBestPerfWMacUJ}{5.37}

\newcommand{\vthreedEOneBestPerfWDramUJ}{36.70}

\newcommand{\vthreedETwoBestPerfWMacUJ}{6.58}

\newcommand{\vthreedEThreeBestPerfWDramUJ}{32.11}

\newcommand{\vthreedEThreeBestPerfWLogicAreaMM}{2.474}

\newcommand{\vthreedXNineBestPerfWDramUJ}{220.20}

%% file: generated/numbers_v3-generalization-design.tex
\newcommand{\vthreegendXFiveBBestPerfWMacPct}{14.1}

%% file: generated/numbers_measurement-validation.tex
\newcommand{\measPeUngatedPJ}{0.640}
\newcommand{\measPeGatedPJ}{0.784}

\newcommand{\measPeGatingPct}{22.6}

\newcommand{\measBaseMeshArrivalNs}{1.668}

\newcommand{\measBaseTileArrivalNs}{13.45}

\newcommand{\measSensGddrPJB}{112.0}

%% file: generated/numbers_v3-dnn512-ppa.tex
\newcommand{\vthreepXOneSBestEdapEdapGain}{159}

\newcommand{\vthreepXOneABestEdapEdapGain}{164}

\newcommand{\vthreepXTwoBestEdapEdapGain}{18.2}

\newcommand{\vthreepXThreeBestEdapEdapGain}{65.4}

\newcommand{\vthreepEOneBestEdapEdapGain}{8.07}

\newcommand{\vthreepETwoBestEdapEdapGain}{20.7}

\newcommand{\vthreepEThreeBestEdapEdapGain}{76.8}

\newcommand{\vthreepXNineBestEdapEdapGain}{1.47}

\newcommand{\vthreepXOneSBestEdapVsEThree}{2.07}
\newcommand{\vthreepXOneABestEdapVsEThree}{2.14}

%% file: generated/numbers_v3-generalization-ppa.tex
\newcommand{\vthreegenpXFiveWBestEdapEdapGain}{54.4}

\newcommand{\vthreegenpXFiveBBestEdapEdapGain}{13.9}

%% file: generated/numbers_v3-workloads.tex
\newcommand{\wlDnnRows}{512}
\newcommand{\wlDnnCols}{512}
\newcommand{\wlDnnDenseCols}{64}
\newcommand{\wlDnnNnz}{8{,}192}
\newcommand{\wlDnnDensityPct}{3.125}

\newcommand{\wlDnnNzBlocks}{512}
\newcommand{\wlDnnTotalBlocks}{1{,}024}

\newcommand{\wlWikiRows}{384}
\newcommand{\wlWikiCols}{384}
\newcommand{\wlWikiDenseCols}{64}
\newcommand{\wlWikiNnz}{3{,}106}
\newcommand{\wlWikiDensityPct}{2.106}

\newcommand{\wlWikiNzBlocks}{489}
\newcommand{\wlWikiTotalBlocks}{576}

\newcommand{\wlObertRows}{256}
\newcommand{\wlObertCols}{512}
\newcommand{\wlObertDenseCols}{64}
\newcommand{\wlObertNnz}{15{,}702}
\newcommand{\wlObertDensityPct}{11.98}

\newcommand{\wlObertNzBlocks}{512}
\newcommand{\wlObertTotalBlocks}{512}
\newcommand{\wlObertNzBlockPct}{100.0}

%% file: generated/numbers_v3-config.tex
\newcommand{\cfgClockFloorNs}{2.0}
\newcommand{\cfgHvRef}{1.1}

\newcommand{\cfgMotpeStartup}{6}

\newcommand{\cfgObertPrunePct}{90}

%% file: generated/numbers_v3-transfer-grid.tex
\newcommand{\vthreexgEThreeOnObertPerfWGain}{3.44}

\newcommand{\vthreexgXOneSOnObertPerfWGain}{1.16}

\newcommand{\vthreexgXOneAOnObertPerfWGain}{1.18}

%% file: generated/numbers_v3-walltime.tex
\newcommand{\wallXOneSMeanIterMin}{20.7}

\newcommand{\wallXOneSLLMSharePct}{10.1}

\newcommand{\wallXOneAMeanIterMin}{21.3}

\newcommand{\wallXOneALLMSharePct}{8.3}